\documentclass[twocolumn,aps,showpacs,prl,10pt]{revtex4-1}
\usepackage{graphicx}
\usepackage{amsmath}
\usepackage{multirow}

\begin{document}

\title{Type-II Dirac semimetals in the YPd$_2$Sn class}

\author{Peng-Jie Guo}
\author{Huan-Cheng Yang}
\author{Kai Liu}\email{kliu@ruc.edu.cn}
\author{Zhong-Yi Lu}\email{zlu@ruc.edu.cn}

\affiliation{Beijing Key Laboratory of Opto-electronic Functional Materials and Micro-nano Devices, Department of Physics, Renmin University of China, Beijing 100872, China}

\date{\today}

\begin{abstract}

The Lorentz-invariance-violating Weyl and Dirac fermions have recently attracted intensive interests as new types of particles beyond high-energy physics, and they demonstrate novel physical phenomena such as angle-dependent chiral anomaly and topological Lifshitz transition. Here we predict the existence of Lorentz-invariance-violating Dirac fermions in the YPd$_2$Sn class of Heusler alloys that emerge at the boundary between the electron-like and hole-like pockets in the Brillouin zone, based on the first-principles electronic structure calculations. In combination  with the fact that this class of  materials was  all reported to be superconductors, the YPd$_2$Sn class provides an appropriate platform for studying exotic physical properties distinguished from conventional Dirac fermions, especially for realizing possible topological superconductivity.

\end{abstract}

\maketitle

Conceptually by generalizing topological characterization from insulator to metal, it has been realized that the Dirac~\cite{a1,2,3,4} and Weyl semimetals~\cite{5,6,7,8,9} are new classes of three dimensional (3D) topological materials, different from the well-studied 3D topological insulators (TI)~\cite{10,11,12,13}. For Dirac semimetals with both time-reversal and space-inversion symmetries, a Dirac point is a linear crossing of two doubly-degenerate bands nearby the Fermi energy with its stability protected by additional symmetry such as a certain crystalline symmetry, while a Weyl point can be obtained by breaking either the time-reversal or the space-inversion symmetry in Dirac semimetals, being a linear crossing of two non-degenerate bands nearby the Fermi energy only available in 3D. A Weyl point acts as a magnetic monopole with either positive or negative charge (chirality) in 3D moment space, while a Dirac point represents a pair of magnetic monopoles with opposite charges. Thus, the Dirac and Weyl semimetals possess a large variety of novel phenomena, such as quantum magnetoresistance~\cite{14} and chiral anomaly~\cite{15}. In addition, the Dirac and Weyl fermions in condensed matter physics correspond to the relativistic Dirac and Weyl fermions in high-energy physics. Nevertheless, the Lorentz invariance is only strictly required  in high-energy physics but not necessary in condensed matter physics ~\cite{16}. This brings hope for condensed matter physicists to discover new types of fermions in real materials.

Very recently, the Lorentz-invariance-violating, namely the type-II, Weyl and Dirac fermions were first conceptually proposed and further predicted in several compounds~\cite{16,17,18,19,20,21,22}, in whose energy spectra a Dirac cone is strongly tilted to a certain direction by the linear kinetic component to break the Lorentz invariance. Experimentally, compounds WTe$_2$ and MoTe$_2$ have been confirmed to host the type-II Weyl fermions~\cite{23,24,25,26,27,28,29,30}, while the type-II Dirac fermions have been verified in PtTe$_2$ by angle-resolved photoemission spectroscopy (ARPES) measurement~\cite{31}. Different from conventional (type-I) Weyl or Dirac fermions, the type-II Weyl or Dirac fermions locate at the boundary between the electron-like and hole-like pockets in the Brillouin zone, expectedly showing many unusual physical properties such as angle-dependent chiral anomaly~\cite{16,32,33} and topological Lifshitz transition~\cite{16,22}.

In comparison with the type-II Weyl semimetals~\cite{23,24,25,26,27,28,29,30}, the type-II Dirac semimetals, which are proposed as improved platforms to afford topological superconductivity and to explore Majorana Fermions~\cite{34}, are much less studied. Up to now, there are only three classes of type-II Dirac semimetals reported: the RbMgBi~\cite{20}, the VAl$_3$~\cite{21}, and the PtSe$_2$ classes~\cite{22}, respectively. The former two are still in theoretical prediction. For the last one, the Dirac points have been experimentally verified but relatively far from the Fermi level~\cite{31}. Moreover, due to their tetragonal or hexagonal crystal structures, these three classes of compounds possess just one pair of Dirac points along the $k_z$ direction in the bulk Brillouin zone. It is thus very important and urgent to search for new type-II Dirac semimetal materials with more pairs of Dirac points sufficiently close to the Fermi level and playing an important role.

The Heusler alloys in real materials are a very large family with abundant physical properties~\cite{35}. Dozens of TIs were theoretically predicted in half Heusler alloys~\cite{36,37,38,39,40}. Among them, some half Heusler alloys are possibly antiferromagnetic topological insulators (AFTI)~\cite{41,42,43}, while some other half Heusler TIs show superconducting behavior~\cite{44,45,46,47} and may realize topological superconductivity~\cite{48,49}. Although the intensive attention has been paid on half Heusler alloys, the likely topological electronic states of full Heusler alloys are rarely studied~\cite{50,51}.

Here, we predict the existence of type-II Dirac fermions in the full Heusler alloys YPd$_2$Sn, ScPd$_2$Sn, ZrPd$_2$Al, HfPd$_2$Al, ZrNi$_2$Al, and HfNi$_2$Al , namely the YPd2Sn class, by  using the  first-principles electronic structure calculations. For this class, we find that there are three pairs of symmetry-protected type-II Dirac fermions located respectively along three equivalent $\Gamma$-X paths in the bulk Brillouin zone. Interestingly, the YPd$_2$Sn class of compounds was all reported to be superconductors in previous experiments ~\cite{52,53,54}. This thus provides a new platform to realize possible topological superconductivity~\cite{34,55}.


The first-principles electronic structure calculations were performed with the projector augmented wave (PAW) method~\cite{56} as implemented in the Vienna ab initio simulation package (VASP)~\cite{57}. The generalized gradient approximation (GGA) of Perdew-Burke-Ernzerhof (PBE) type was adopted for the exchange-correlation potential. The kinetic energy cutoff of the plane wave basis was set to be 340 eV. A 10$\times$10$\times$10 $\textbf{k}$-point mesh for the Brillouin zone (BZ) sampling and the Gaussian smearing with a width of 0.05 eV around the Fermi surface were adopted. Both cell parameters and internal atomic positions were fully relaxed until the forces on all atoms were smaller than 0.01~eV/{\AA}.
The maximally localized Wannier functions (MLWF) \cite{58,59} method was used to calculate the Fermi surface. To study the topological surface states, a 1$\times$1 two-dimensional supercell with an AB$_2$C slab containing 66 atoms and a 15 {\AA} vacuum was employed to simulate the AB$_2$C (001) surface.


\begin{figure}[!t]
\includegraphics[angle=0,scale=0.12]{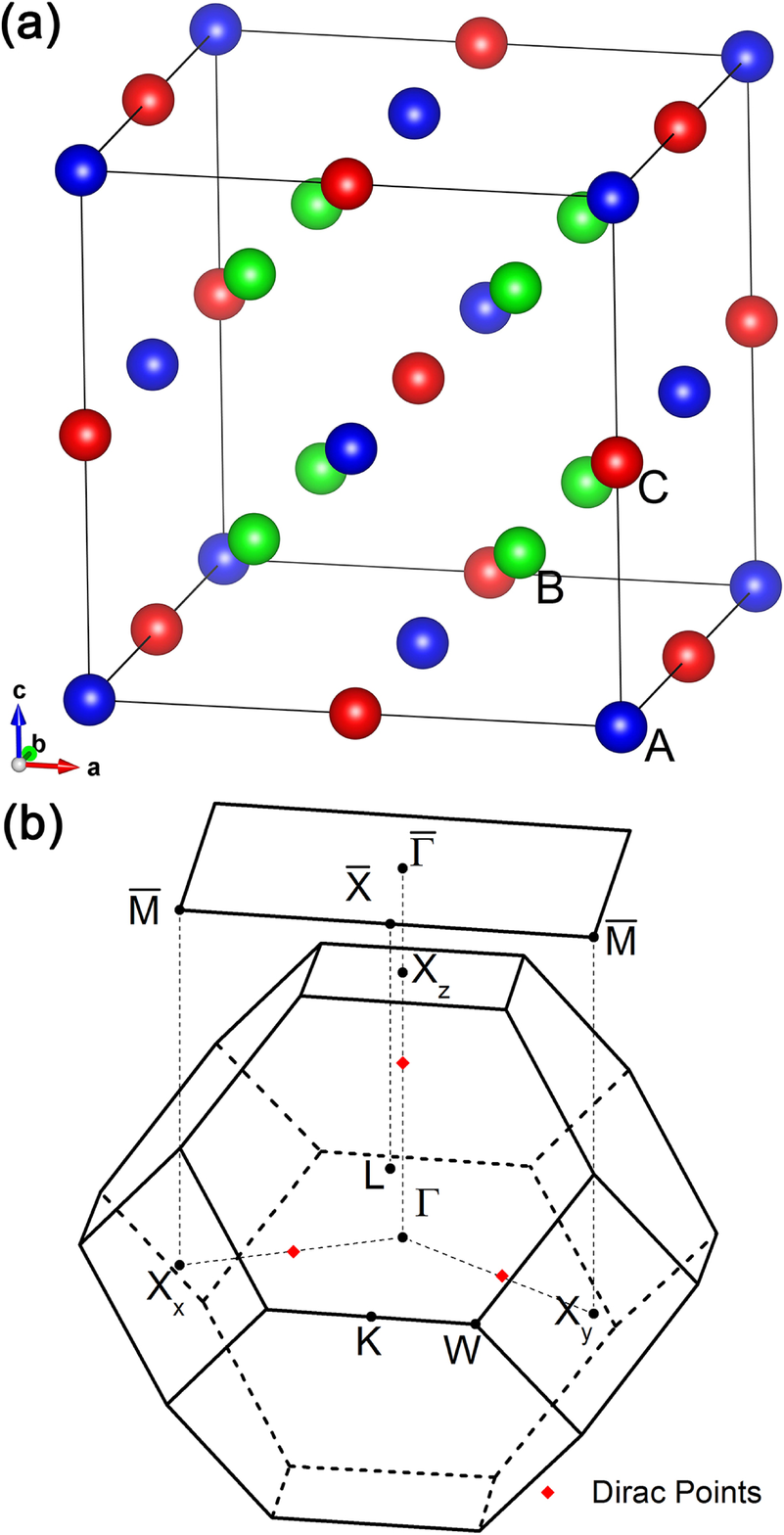}
\caption{(Color online) (a) Crystal structure of the full Heusler alloys AB$_2$C with the Fm\=3m symmetry. (b) Bulk Brillouin zone and the projected surface Brillouin zone of an AB$_2$C(001) surface. The black dots represent the high-symmetry $\textbf{k}$ points in Brillouin zone while the red ones denote the Dirac points.}
\label{Fig1}
\end{figure}

\begin{figure}[!t]
\includegraphics[angle=0,scale=0.44]{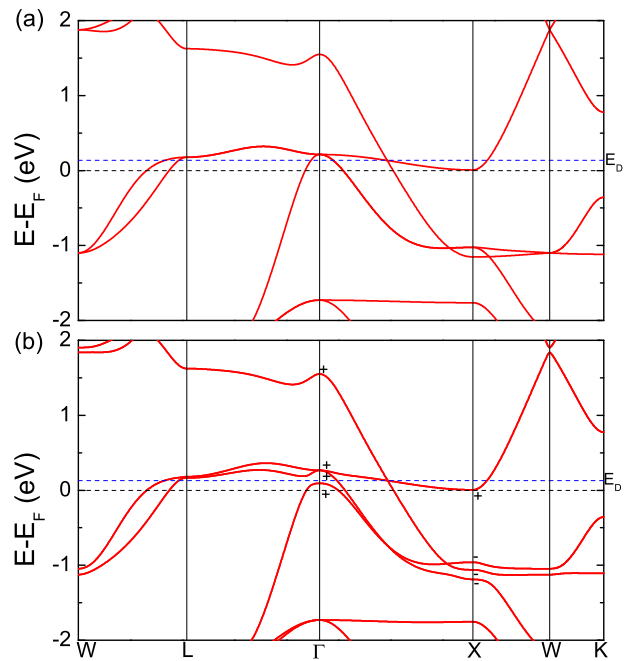}
\caption{(Color online) Band structures along high-symmetry directions of the Brillouin zone for YPd$_2$Sn calculated (a) without and (b) with spin-orbital-coupling (SOC) effect. The Fermi level is set to zero. The $E_\text{D}$ denotes the energy of the Dirac point. The signs '$+$' and '$-$' indicate the parities of corresponding states at the high-symmetry points $\Gamma$ and X.}
\label{Fig2}
\end{figure}

\begin{table}[b!]
\caption{\label{tab:I} Calculated and experimental (in parentheses) lattice constants $a$ (in unit of {\AA}) of several full Heusler alloys. The locations of Dirac points along the $\Gamma$-$X$ directions in the BZ ($k_x$) and their energies ($E_\text{D}$) with respect to the Fermi level. The superconducting transition temperatures $T_c$s reported in literatures.}
\begin{center}
\begin{tabular*}{1.0\columnwidth}{@{\extracolsep{\fill}}ccccccc}
\hline\hline
 &      $a$(\AA) &      $k_x$($\pi/a$)&      $E_\text{D}$(meV)  &    $T_c$(K)\\
\hline
  ZrNi$_2$Al &	6.122(6.106$^{a}$) &  0.381 & -296 & 1.38$^{a}$  \\
  HfNi$_2$Al &  6.088(6.069$^{a}$) &  0.361 & -257 & 0.74$^{a}$  \\
  ZrPd$_2$Al &  6.451(6.390$^{b}$) &  0.363 & -270 & 3.2$^{b}$   \\
  HfPd$_2$Al &  6.417(6.370$^{b}$) &  0.348 & -232 & 3.4$^{b}$   \\
  ScPd$_2$Sn &  6.580(6.509$^{c}$) &  0.239 &  142 & 2.25$^{c}$  \\
  YPd$_2$Sn  &  6.787(6.718$^{a}$) &  0.223 &  129 & 4.9$^{a}$   \\
\hline\hline
$^{a}$Ref.~\cite{52}.\\
$^{b}$Ref.~\cite{53}.\\
$^{c}$Ref.~\cite{54}.
\end{tabular*}
\end{center}
\end{table}

The crystal structure of the full Heusler alloys AB$_2$C takes the Fm\=3m space group symmetry [Fig. 1(a)]. The elements A and C make up the rock salt structure, and the element B locates at the ($\frac{1}{4}$, $\frac{1}{4}$, $\frac{1}{4}$) coordinate and its equivalent positions. As listed in Table I, the fully relaxed lattice constants of the Heusler alloys from our calculations accord well with their corresponding experimental values~\cite{52,53,54}. In the bulk BZ of the Heusler alloys we studied [Fig. 1(b)], the black dots represent the high-symmetry $\textbf{k}$ points, while the red dots schematically display the locations of Dirac points along the $\Gamma$-X directions.

We first show the electronic band structure of YPd$_2$Sn as a representative case of the Dirac semimetals in the full Heusler alloys we predict (Fig. 2). Without spin-orbital coupling (SOC), a band crossing close to the Fermi level along the $\Gamma$-X direction can be clearly observed [Fig. 2 (a)]. Once the SOC is included in the calculations, two doubly degenerate bands along the $\Gamma$-L and $\Gamma$-X directions split [Fig. 2(b)]. Nevertheless, the above band crossing along the $\Gamma$-X direction still holds with only a minor energy shift. As these two bands are both doubly-degenerate bands due to the time-reversal and the space-inversion symmetries in YPd$_2$Sn, this crossing point is fourfold degenerate. We thus suggest that this band crossing point is a Dirac point protected by the C$_{4v}$ double group in YPd$_2$Sn. To prove it, we have calculated the parities for the bands around the Fermi level at high-symmetry points $\Gamma$ and X and analyzed the orbital characteristics of these two crossing bands.

\begin{figure}[!t]
\includegraphics[angle=0,scale=0.14]{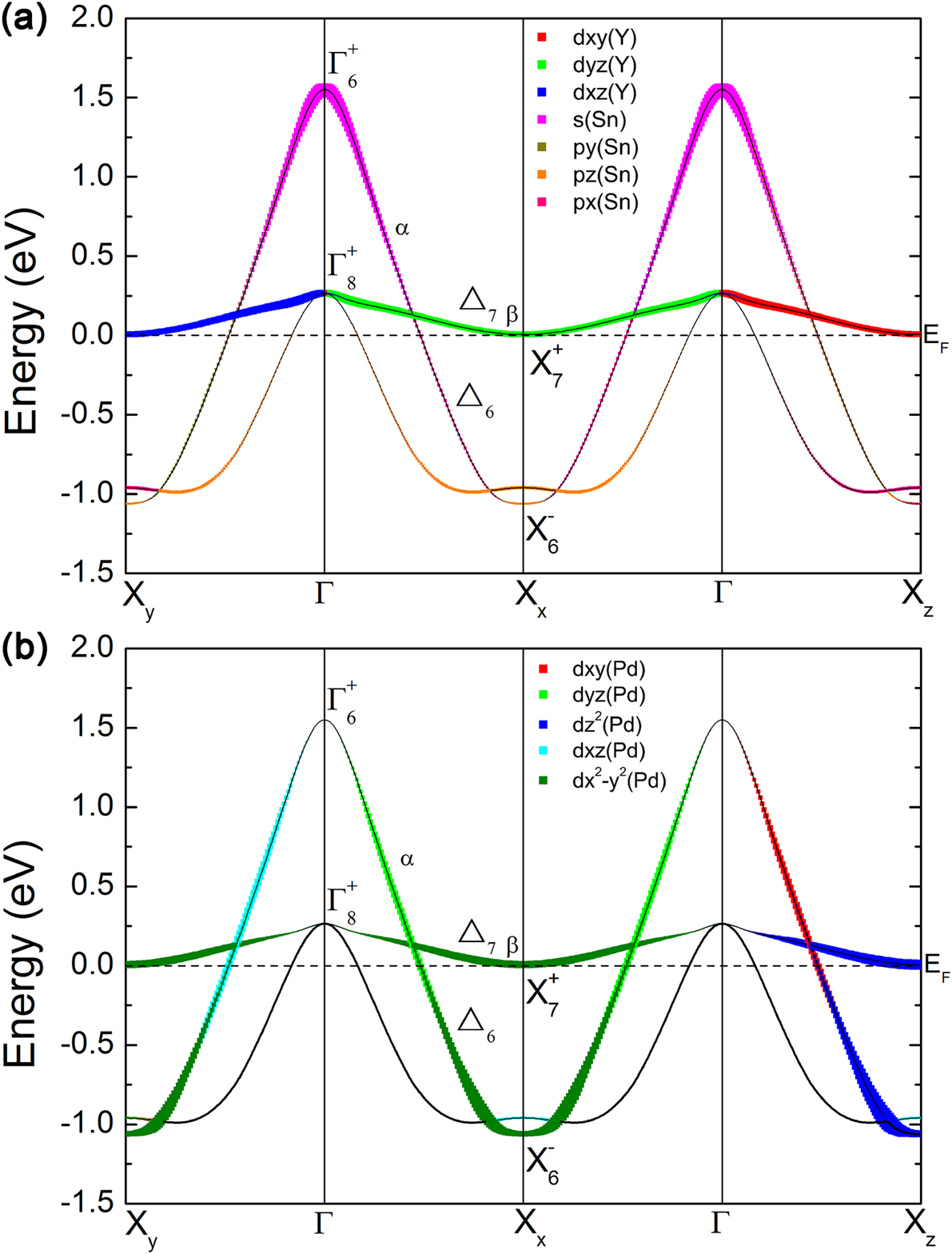}
\caption{(Color online) Orbital characteristics of the two crossing bands $\alpha$ and $\beta$ contributed from different orbitals of (a) Y/Sn and (b) Pd atoms and their symmetries along the $\Gamma$-X directions. The Fermi level is set to zero. The signs '$+$' and '$-$' indicate the parities of corresponding states.}
\label{Fig3}
\end{figure}

As there are three equivalent X points in the BZ for the cubic symmetry [Fig. 1(b)], we have studied the orbital information of the crossing bands along three different $\Gamma$-X directions. By analyzing all orbitals ($s, p$, and $d$) of Y, Pd, and Sn atoms, we find that the two crossing bands ($\alpha$ and $\beta$) mainly consist of the 4$d$ orbitals of Y atom, the 5$s$ and 5$p$ orbitals of Sn atom, and the 4$d$ orbitals of two centrosymmetric Pd atoms (Fig. 3). Among them, the 5$s$ orbital of Sn atom has isotropic contribution to the two crossing bands along three different $\Gamma$-X directions. In contrast, the contributions from the 5$p$ and 4$d$ orbitals are anisotropic.

For the $\alpha$ band, the orbital characteristics change gradually from Sn $s$ orbital at the $\Gamma$ point to Sn $p$ orbital [Fig. 3(a)] as well as the $d$ orbitals of two centrosymmetric Pd atoms toward the X point [Fig. 3(b)]. By analyzing the parity and the orbital characteristics, the symmetries are $\Gamma^{+}_{6}$ at the $\Gamma$ point, $\Delta_{6}$ along the $\Gamma$-X axis, and X$^{-}_{6}$ at the X point, respectively. On the other hand, for the $\beta$ band, the contributions change from Y $t_{2g}$ orbitals at the $\Gamma$ point to both the $t_{2g}$ orbitals of Y atom and the $e_{g}$ orbitals of two centrosymmetric Pd atoms toward the X point. The respective symmetries are $\Gamma^{+}_{8}$, $\Delta_{7}$, and X$^{+}_{7}$ at the $\Gamma$ point, along the $\Gamma$-X direction, and at the X point. Here $\Delta_{6}$ and $\Delta_{7}$ are respectively two inequivalent irreducible representations of the C$_{4v}$ double group in YPd$_2$Sn, therefore these crossing points of the $\alpha$ and $\beta$ bands in the $\Gamma$-X directions are protected by the symmetry of the C$_{4v}$ double group.

\begin{figure}[!t]
\includegraphics[angle=0,scale=0.26]{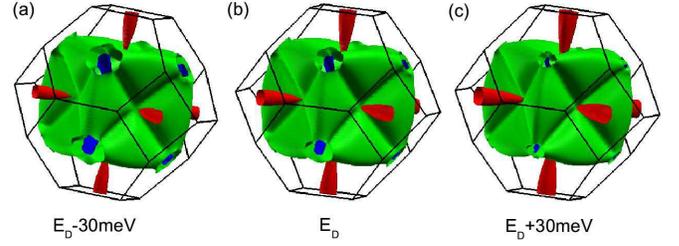}
\caption{(Color online) The isoenergetic surfaces of YPd$_2$Sn at (a) $E_\text{D}$-30meV, (b) $E_\text{D}$, and (c) $E_\text{D}$+30meV in bulk BZ. The red surfaces are electron-like pockets while the green and blue surfaces are hole-like pockets. Contact between the electron-like and hole-like pockets occurs at the Dirac points when $E$=$E_\text{D}$.}
\label{Fig4}
\end{figure}

More importantly, the Dirac cone is tilted strongly along the $\Gamma$-X axis, which is a typical feature of the type-II Dirac fermions. According to our calculations on YPd$_2$Sn, there are three pairs of symmetry-protected Dirac points, which locate at $k$ = (0, 0, $\pm$0.223), (0, $\pm$0.223, 0), and ($\pm$0.223, 0, 0) (in unit of $\pi/a$) on the equivalent $\Gamma$-X axes with an energy of $E_\text{D}$ = 129 meV above the Fermi level. Remarkably, the electronic band structures of other YPd$_2$Sn class materials (ZrNi$_2$Al, HfNi$_2$Al, ZrPd$_2$Al, HfPd$_2$Al, and ScPd$_2$Sn) are very similar (see the Supplemental Materials). The respective positions of the Dirac points for this class of materials are summarized in Table I. Among them, the $E_\text{D}$ of YPd$_2$Sn is the closest to the Fermi level.

Based on the above parity, orbital-characteristics, and group-theory analysis of the crossing bands $\alpha$ and $\beta$, we clarify that the type-II Dirac points exist in the electronic  band structure of YPd$_2$Sn. For the type-I Dirac semimetals like Na$_3$Bi and Cd$_3$As$_2$~\cite{3,4}, the Fermi surface becomes a number of points when the Fermi level locates at the Dirac-point energy ($E_\text{D}$). In contrast, for the type-II Dirac semimetal, a part of the upper Dirac cone is lower than a part of the lower cone, thus retaining both electron-like and hole-like pockets when the Fermi level crosses $E_\text{D}$. This is exactly the case for YPd$_2$Sn as shown in Fig. 4(b). Furthermore, when the Fermi level is lower or higher than $E_\text{D}$, the electron-like pocket and hole-like pocket will separate from each other as demonstrated in Figs. 4(a) and (c). Therefore, a Lifshitz transition of the Fermi surfaces can be observed in the type-II Dirac semimetals by tuning the Fermi level.

\begin{figure}[!t]
\includegraphics[angle=0,scale=0.18]{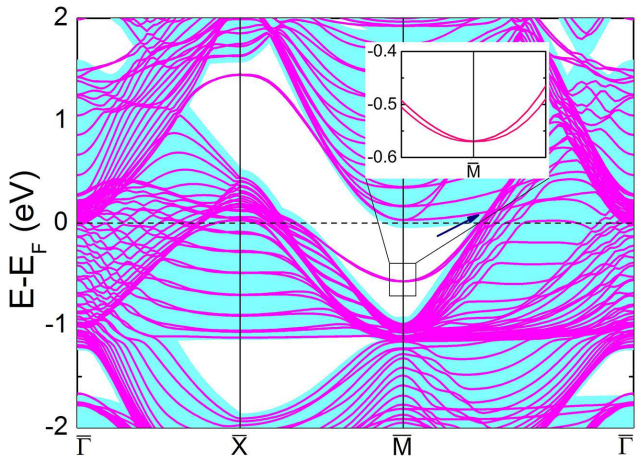}
\caption{(Color online) Surface band structure of the YSn-terminated YPd$_2$Sn (001) surface along the high symmetry directions of the surface Brillouin zone [Fig. 1(b)]. The shaded cyan areas denote the surface-projected bulk states.}
\label{Fig5}
\end{figure}

The symmetry-protected topological surface states are one of the most important phenomena for the Dirac semimetals. We thus have calculated the surface states of YPd$_2$Sn (Fig. 5). For the YPd$_2$Sn (001) surface, two Dirac points along the $\Gamma$-X$_z$ direction in the bulk BZ are projected  to the $\overline{\Gamma}$ point in the surface BZ and the other four Dirac points are projected to the $\overline{\Gamma}$-$\overline{\text{M}}$ directions of the surface BZ [Fig. 1(b)]. As shown in Fig. 5, two surface states below the Fermi level around $\overline{\text{M}}$ merge into the bulk continuum at the projected bulk Dirac point along the $\overline{\Gamma}$-$\overline{\text{M}}$ direction (marked by the blue arrow). These are the symmetry-protected topological surface states~\cite{20}. The other surface states, for instance the surface states above the Fermi level at $\overline{\text{X}}$ in the surface BZ, do not conflate with the bulk continuum at the bulk Dirac point. We thus deduce that the other surface states are normal surface states.




The type-II Dirac semimetals predicted in this study have many advantages. Firstly, due to the Fm\=3m symmetry of their face-centered-cubic (fcc) primitive cells (Fig. 1), there are more Dirac-point pairs in the bulk BZ of the YPd$_2$Sn class than those of the hexagonal PtSe$_2$ class as well as the tetragonal RbMgBi and VAl$_3$ families. These Dirac points remain after the surface projection and their corresponding topological surface states are prone to be detected in ARPES measurement. Secondly, 
the Dirac points of the YPd$_2$Sn class are all close to the Fermi level (Table I). Especially, the Dirac-point energy of YPd$_2$Sn is just 129 meV above the Fermi level, which reduces to 75 meV when the lattice constants are compressed by 3\%. By tuning the Fermi level with a gate voltage, the angle-dependent chiral anomaly may be observed in transport experiment.
Last but not least, in different YPd$_2$Sn class materials, the Dirac points locate either above or below the Fermi level (Table I). Thus by proper element substitution or lattice modulation, the energy of a Dirac point may well locate on the Fermi level.

Another striking point is that the YPd$_2$Sn class of compounds was all reported to be superconductors in previous experiments (Table I). As suggested by a recent study \cite{34}, compared with the Weyl semimetals and the conventional Dirac semimetals, a wealth of density of states provided by the electron-like and hole-like pockets near the Dirac points in the type-II Dirac semimetals are benefitial to the carrier ratio of a possible topological superconductor. Thus, the YPd$_2$Sn class may serve as prototypical systems to study topological superconductivity and to search for Majorana fermions, which have potential applications in quantum computation. Furthermore, from the crystal-structure standpoint, the YPd$_2$Sn class belongs to the natural abundant Heusler compounds with versatile nonmagnetic, magnetic, semiconducting, metallic, superconducting, and topological properties~\cite{35}. We expect that the full (half) Heusler compounds will provide plenty of chances to discover new topological semimetals and possible topological superconductors.



By using the first-principles electronic structure calculations, we predict the existence of type-II Dirac fermions that locate at the boundary between the electron-like and hole-like pockets in the YPd$_2$Sn class. Benefited from the symmetry of their fcc crystal structures, three pairs of type-II Dirac points appear on the $\Gamma$-X directions of the bulk BZ. Considering that these compounds were all reported to be superconductors, the YPd$_2$Sn class provides a new platform to study exotic physical properties distinguished from conventional Dirac fermions and to realize possible topological superconductivity.


We thank Yuan-Yao He, Xin-Zheng Li, Zheng-Xin Liu, Ning-Hua Tong, Rui Lou, and Tao Li for helpful conversations. This work was supported by the National Natural Science Foundation of China (Grants No. 11474356 and No. 91421304), the Fundamental Research Funds for the Central Universities, and the Research Funds of Renmin University of China (Grants No. 14XNLQ03 and No. 16XNLQ01). Computational resources have been provided by the Physical Laboratory of High Performance Computing at Renmin University of China. The Fermi surfaces were prepared with the XCRYSDEN program ~\cite{60}.


\end{document}